\documentclass[12pt]{iopart}
\usepackage{graphicx}
\begin{document}


\title{The dual nature of 5f electrons and origin of heavy fermions in
U compounds}

\author{G Zwicknagl\dag and P Fulde\ddag}

\address{\dag\ Institut f\"ur Mathematische Physik, Technische Universit\"at 
Braunschweig, Mendelssohnstr. 3, 38106 Braunschweig, Germany}
\address{\ddag\ Max-Planck-Institut f\"ur Physik komplexer Systeme,
N\"othnitzer Str. 38, 01187 Dresden, Germany}

\date{\today}

\begin{abstract}
We develop a theory for the electronic excitations in UPt$_3$ which is based on
the localization of two of the $5f$ electrons. The remaining $f$ electron is
delocalized and acquires a large effective mass by inducing intra-atomic
excitations of the localized ones. The measured deHaas-vanAlphen frequencies
of the heavy quasiparticles are explained as well as their anisotropic heavy
mass. A model calculation for a small cluster reveals why only the largest of
the different $5f$ hopping matrix elements is operative causing the electrons
in other orbitals to localize.
\end{abstract}




\section{Introduction}
There is growing evidence that actinide ions may have localized 
as well as delocalized $ 5f$ electrons. This picture which was
suggested by transport measurements  
\cite{Schoenes96} is supported by a great variety of experiments
including, e.~g., photoemission and neutron inelastic scattering 
experiments on 
UPd$ _{2} $Al$ _{3} $ \cite{Takahashi95,Metoki98,Bernhoeft98} as well
as muon spin 
relaxation measurement in UGe$ _{2} $ \cite{Yaouanc02}.
The assumption is further supported
by quantum chemical calculations on uranocene U(C$ _{8} $H$ _{8} $)$ _{2} $
\cite{Liu97} which exhibit a number of low-lying excitations which are due to
rearrangements of the 5$ f $ electrons. There are speculations that
the presence of localized $ 5f $-states might
even be responsible for the attractive interactions leading to superconductivity
\cite{Sato01}. We should like to mention that the dual model should allow 
for a rather natural description of heavy fermion superconductivity 
coexisting with 5f-derived magnetism.

The above-mentioned observations form the basis of the dual model which 
provides a microscopic theory for the heavy quasiparticles in
U compounds. The ansatz conjectures that the delocalized 5 $f$ states
hybridize with the conduction states and form energy bands while the
localized ones form multiplets to reduce the local Coulomb repulsion. The two
subsystems interact which leads to the mass enhancement of the delocalized
quasiparticles. The situation resembles that in Pr metal where a mass 
enhancement of the conduction electrons by a factor of 5 results from 
virtual crystal field (CEF) excitations of localized 4 $ f^{2} $ 
electrons \cite{White81}.

The dual ansatz reproduces the dHvA data in 
UPt$ _{3} $ \cite{Zwicknagl02} and UPd$ _{2} $Al$ _{3} $. Detailed
Fermi surface studies for UGa$ _{3} $
\cite{Biasini02} and high-resolution photoemission measurements for
URu$_2$Si$_2$  
\cite{Denlinger01} show that the observed Fermi surfaces cannot be
explained by assuming all 5 $f$ electrons to be itinerant or localized. 
Measurements of the optical conductivity in UPd$ _{2} $Al$ _{3} $ and
UPt$ _{3}$ \cite{Dressel02} indicate that the enhanced effective masses
m$^*$ of the quasiparticles should result from the interaction of
delocalized states with localized magnetic moments. 

The coexistence of itinerant and localized 5 $f$ states is referred to 
as partial localization. It plays an important role in many
intermetallic actinide compounds. Partial localization arises from 
interplay between the hybridization of the 5 $f$ states with the 
conduction electrons and the local Coulomb
correlations. The underlying microscopic mechanism is an
area of active current research \cite{Lundin00,Soederlind00}. LDA calculations
show that the hopping matrix elements for different $5f$ orbitals vary. But it
is of interest to understand why only the largest one of them is important and
why the other ones are suppressed.

In order to justify the above assumption we present
model calculations which focus on the interplay between 
delocalization of 5 $f$ states and Hund's rule correlations.
The results \cite{Efremov02} clearly show how partial localization can 
arise in 5 $f$ systems. In addition, they suggest rather complex phase 
diagrams depending upon the strengths of the hopping matrix referring to
different orbital
elements. Variation of the intersite hopping by applying (hydrostatic) pressure
should lead to new types of (quantum) phase transitions. We think the 
present model could be used to study the pressure dependence of the
magnetization in UGe$_2$.

In addition to the full model Hamiltonian, we investigate a
simplified version which treats the local Coulomb interaction in close
analogy to LDA+U. The results allow us to assess the general validity
of this popular approximation scheme. Finally we compare the results
of the full model Hamiltonian with those obtained from a Hartree-Fock
approximation.

\section{Heavy quasiparticles in UPt$ _{3} $ and 
UPd$ _{2} $Al$ _{3}$: Dual model}
\begin{figure}[h t b]
\vspace{5mm}
\begin{center}
\begin{minipage}[t]{17cm}
\begin{minipage}[b]{80mm}
\begin{center}
{\Large\bf UPt$_3$}
\end{center}
\includegraphics[width=75mm]{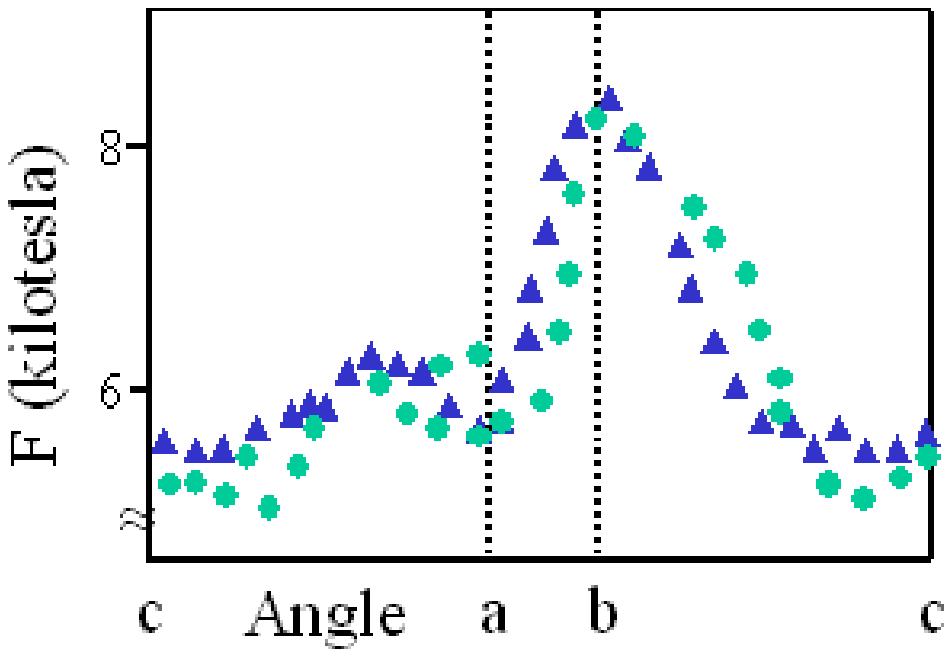}
\end{minipage}
\hfill
\begin{minipage}[b]{80mm}
\begin{center}
{\Large\bf }
\end{center}
\includegraphics[width=55mm]{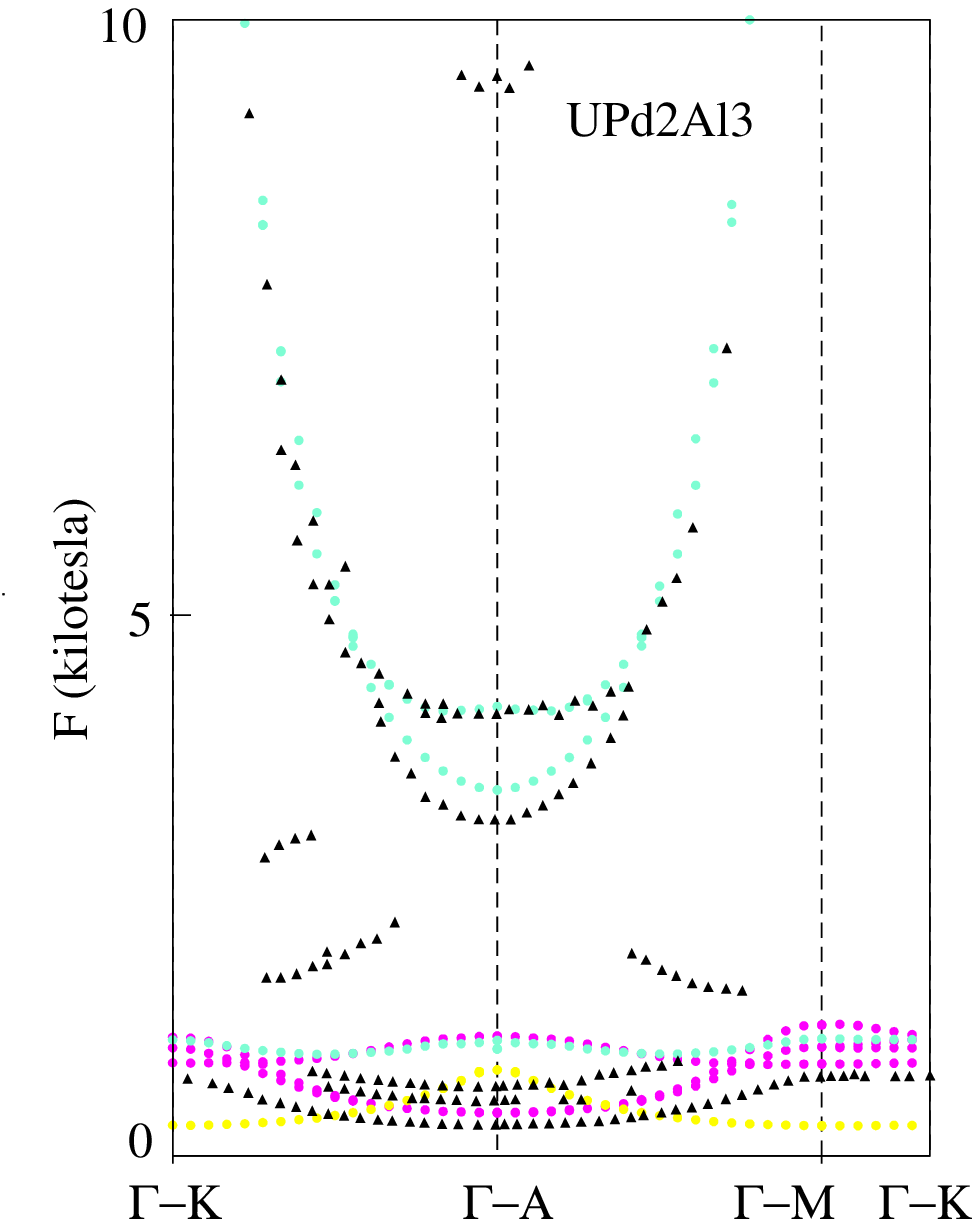}
\end{minipage}
\end{minipage}
\end{center}
\caption{DeHaas-vanAlphen cross sections for the heavy 
quasiparticles as calculated within the 
dual model (circles) \cite{Zwicknagl02}. The experimental data
for UPt$_3$ (triangles)
are from \cite{Kimura98} while those for UPd$_2$Al$_3$ (black
triangles) are taken from \cite{Inada99}}
\label{fig:HeavyFermiSurfaceCrossSections}
\end{figure}

We calculate the heavy quasiparticles in UPt$_3$ and UPd$_2$Al$_3$ 
within the dual model 
considering two of the 5$ f $ electrons as localized, in agreement
with the absence of any Kramers doublets in cases where a crystalline electric
field (CEF) splitting
of U states has been observed. The calculation of the heavy bands proceeds in
three steps as described in \cite{Zwicknagl02}: First, the
band-structure is determined  by solving
the Dirac equation for the self-consistent LDA potentials 
but excluding the U 5$ f $~ j=$ \frac{5}{2} $, j$ _{z} $=$ \pm \frac{5}{2} $
and j$ _{z} $=$ \pm \frac{1}{2} $ states from forming bands. The localized
5$ f $ orbitals are accounted for in the self-consistent density and, 
concomitantly, in the potential seen by the conduction electrons. The 
intrinsic bandwidth of the itinerant U 5$ f $ j=$ \frac{5}{2} $, 
j$ _{z} $=$ \pm \frac{3}{2} $ electrons is taken from the LDA
calculation while the position of the corresponding
band-center C is chosen such that the density distribution of the conduction
states as obtained within LDA remains unchanged.The $ f $ occupancy per U
atom for the delocalized 5$ f $ electrons amounts to n$ _{\textrm{f}} $
= 0.65 indicating that we are dealing with a mixed valent situation. 
The calculated dH-vA frequencies agree rather well with the observed
ones \cite{Kimura98} as shown in Figure 
\ref{fig:HeavyFermiSurfaceCrossSections}. We also include the
corresponding results for UPd$_2$Al$_3$ which are compared to
experimental data from \cite{Inada99}.

In the second step, the localized U 5$ f $ states are calculated 
assuming the jj-coupling scheme. The
Coulomb matrix elements are calculated from the radial functions of
the ab-initio band structure potentials. We find a doubly degenerate ground
state with J$ _z$ = $\pm 3 $ which must be an eigenstate of J = 4 and
has an overlap of 0.865 with the Hund's
rule ground state $ ^{3}H_{4} $ derived from the LS-coupling
scheme. In the hexagonal symmetry, the two-fold degeneracy of the 
ground-state is lifted by a CEF yielding the two
states \hbox{$|\Gamma_{3} \rangle$} and \hbox{$| \Gamma_{4} \rangle$}. Note that $ |\Gamma _{4}\rangle  $ has
been suggested as ground state of UPd$ _{2} $Al$ _{3} $. We assume that
the splitting energy $ {\tilde{\delta }} $ between $ |\Gamma _{4}\rangle  $
and $ |\Gamma _{3}\rangle  $ is of order 20 meV for UPt$_3$.  
The coupling between the localized and delocalized $ f $ electrons is 
directly obtained from the expectation
values of the Coulomb interaction U$ _{\textrm{Coul}} $ in the 5$ f^{3} $
states $ M=\langle f^{1};\frac{5}{2},\frac{3}{2}|\otimes \langle \Gamma _{4}|U_{\textrm{Coul}}|\Gamma _{3}\rangle \otimes |f^{1};\frac{5}{2},\frac{3}{2}\rangle =0.19eV $.

Finally, we determine the renormalization of the effective masses
which results from the coupling between the two $ 5f $ subsystems. 
\begin{table}
\caption{Measured and calculated effective masses of UPt$_3$ 
for various directions of the magnetic field}
\label{tab:EffMassesUPt3}
\begin{center}
\begin{tabular}{lccc}
\hline 
$ m^{*} $&
 c &
 a &
 b \\
\hline 
Experiment &
 110 &
 82 &
 94 \\
 Theory &
 128 &
 79 &
 104  \\
\hline 
\end{tabular}
\end{center}
\end{table}
The enhancement factor is calculated from the self-consistent
solution of the self-energy equation \cite{White81} with the input
taken from ab-initio electronic structure calculations for the
delocalized and the localized 5 $f$ electrons. The density of states with
two localized 5 $f$ electrons is N(0) $\simeq$ 15.5 states/(eV
cell), the 5 $f$-weight per spin and U atom of the band states amounts 
to 4a$^2$=0.13 while the transition matrix element between the
low-lying singlet states in the localized 5 $f^2$ shell in the
presence of a CEF equals $|$M$|^2$=0.036 e V$^2$. The only adjusted
parameter is the energy ${\tilde \delta}$ characterizing the centers
of gravity of the CEF excitations. By comparison with other U compounds such as
UPd$_2$Al$_3$ we estimate ${\tilde \delta} \simeq$ 20 meV.
This general concept reproduces the quasiparticles rather well 
as can be seen from the 
results summarized in Table \ref{tab:EffMassesUPt3}. A similar analysis applies for UPd$_2$Al$_3$.

\section{Partial localization from competition between angular
correlations and hopping}

The calculations start from small clusters which model the U sites in heavy
fermion compounds. We keep only the degrees of freedom of the 5f shells the
conduction states being accounted for by (effective) anisotropic
intersite hopping. Here we consider the two-site model. The general
results qualitatively agree with those found for a three-site
cluster. The Hamiltonian reads
\begin{equation}
H  =  \sum _{j_{z}}\,
t_{j_{z}}\left(c_{j_{z}}^{\dagger}(1)c_{j_{z}}(2)+h.c.\right) + H_{Coul} 
\label{eq:TwoSiteHamiltonian}
\end{equation}
Here $c_{j_{z}}^{\dagger}(i)$ ($c_{j_{z}}(i)$), create (annihilate) an
electron at site $i$ (=1,2) in the $5f\ j=5/2$ state with
$j_z=-5/2,\dots,5/2$. The effective hopping between the sites is
chosen to be diagonal in the orbital index $ j_{z}$ which seems to be
compatible with 
LDA calculations for the U-based heavy-fermion systems
UPt$ _{3} $ and UPd$ _{2} $Al$ _{3} $.
The local Coulomb repulsion 
\begin{equation}
H_{Coul}= 
\sum_{i=1,2}\ \sum_{j_{z1}>j_{z2}}\sum_{j_{z3}>j_{z4}}
\langle j_{z1}j_{z2}\left| U\right| j_{z3}j_{z4}\rangle
c_{j_{z1}}^{\dagger}(i)c_{j_{z2}}^{\dagger}(i)c_{j_{z3}}(i)c_{j_{z4}}(i) 
\end{equation}
depends upon the Coulomb matrix elements 
\begin{equation}
\langle j_{z1}j_{z2}| U | j_{z3}j_{z4}\rangle = 
\delta _{j_{z1}+j_{z2},j_{z3}+j_{z4}} \sum_J \langle 5/2j_{z1}5/2j_{z2} |
JJ_{z}\rangle U_{J}\langle JJ_{z} | 5/2j_{z3}5/2j_{z4}\rangle 
\end{equation}
 where $J$ denotes the total angular momentum and 
$J_{z}=j_{z1}+j_{z2}=j_{z3}+j_{z4} $. The sum is restricted to even
values of $ J $, i. e., $ J=0,2,4 $ to satisfy the antisymmetry requirement.
In the actual calculations, we use the parameters $ U_{J} $ determined from
the LDA 5f wave functions in UPt$ _{3} $, i.~e.,~$U_{J=4}=17.21 eV$, 
$U_{J=2}=18.28 eV$, and $U_{J=0}=21.00 eV$. Finally, the 
$ \left\langle 5/2j_{z1}5/2j_{z2}\left. \right| JJ_{z}\right\rangle  $
denote the Clebsch-Gordan coefficients.

To simulate the situation in the U-based heavy-fermion compounds we consider
the model in the intermediate valence regime with an average f-valence
close to 2.5. 

The eigenstates of the Hamiltonian Eq. (\ref{eq:TwoSiteHamiltonian})
are characterized by 
 $J_z=J_z^{(1)}+J_z^{(2)}$ where $J_z$ is the z-component of the total 
angular momentum of the two-site system while
$J_z^{(1)}$ and $J_z^{(2)}$ refer to the angular momentum projections
of the individual sites. 
\begin{figure}[h t b]
\begin{center}
\includegraphics[width=0.6\columnwidth,clip]{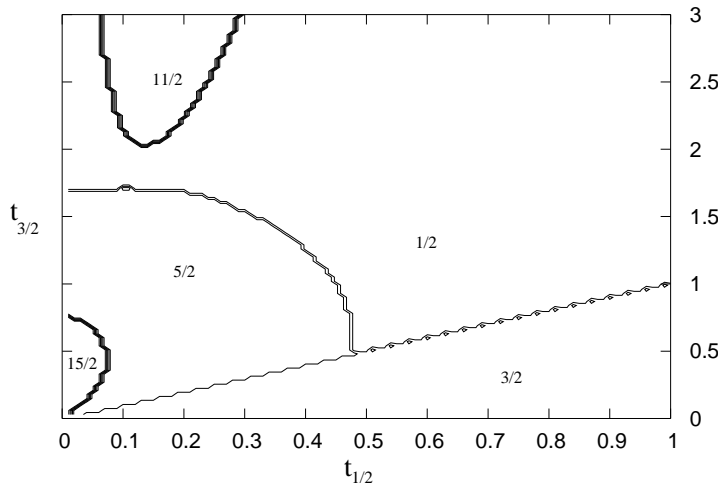}
\end{center}
\caption{
Magnetization of the ground state. The model
predicts two ``high-spin'' phases ($J_z=15/2$ 
and $J_z=11/2$) with  ferromagnetic intersite 
correlations $\langle \vec{J}^{(1)} \cdot \vec{J}^{(2)} \rangle$ 
for strong anisotropy 
$t_{3/2} \gg t_{1/2}=t_{5/2}$ and three 
``low-spin'' phases  $J_z=5/2,1/2,5/2$, 
respectively.}
\label{fig:JzRestrictedModel}
\end{figure}
We study the evolution of the ground state with the hopping
parameters  $t_{3/2}$, $t_{5/2}$,and $t_{1/2}$. The energy variation
is smooth except for a kink along 
the isotropic line $t_{1/2}=t_{3/2}=t_{5/2}$. The character of the
ground state, however, changes as can be seen by considering the total 
magnetization $J_z$ of the ground state displayed in Figure
\ref{fig:JzRestrictedModel}. The phase diagram is strongly affected by 
magnetic fields.
Standard electronic structure calculations for extended systems
such as the Hartree-Fock method or an LDA+U-type ansatz
generally overestimate the stability of the ferromagnetic phases
and fail to describe the subtle breaking-up of the Hund's rule
correlations. 
\begin{figure}[h t b]
\begin{center}\begin{minipage}[t]{17cm}
\begin{minipage}[t]{80mm}
\includegraphics[width=75mm]{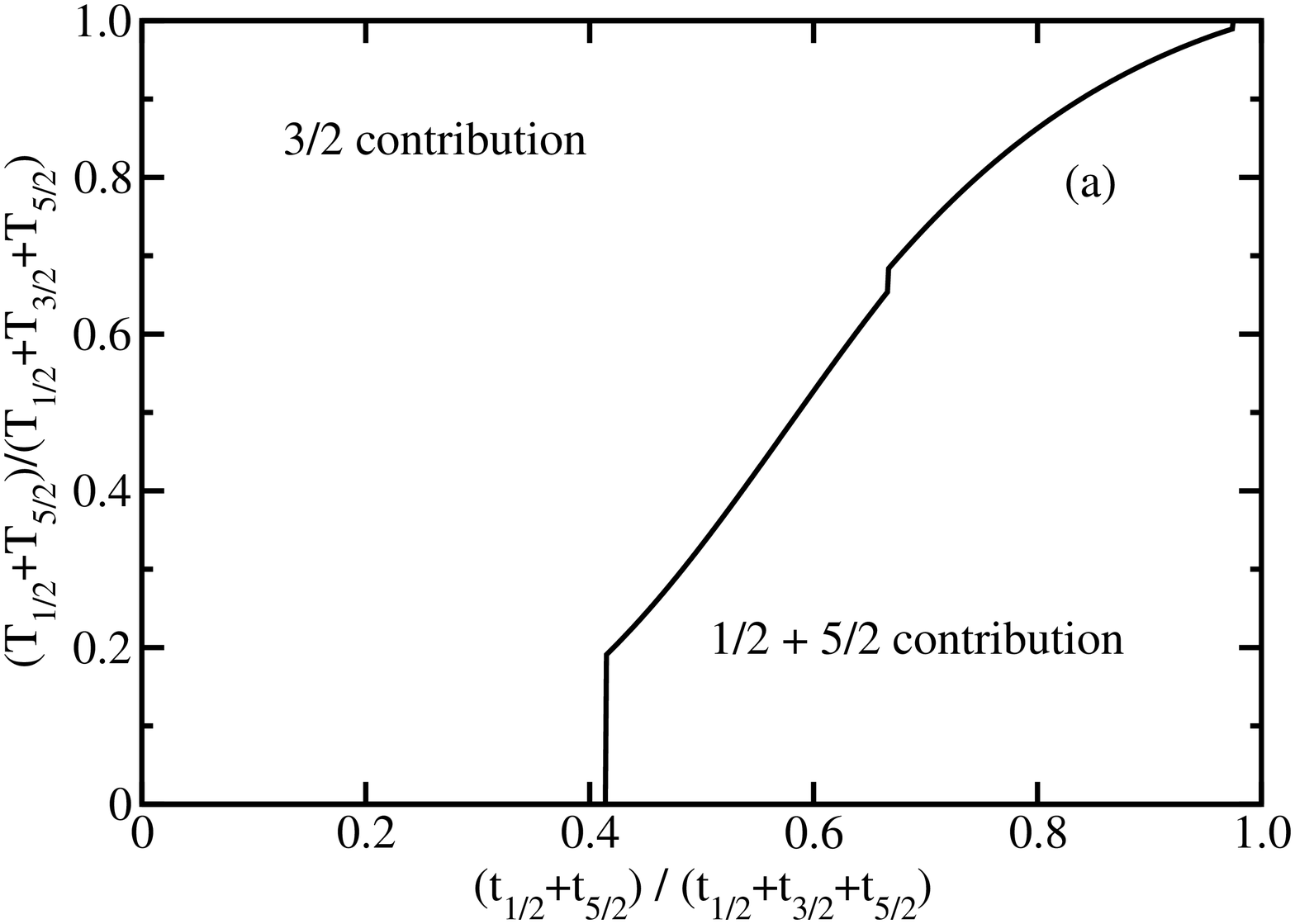}
\end{minipage}
\hfill
\begin{minipage}[t]{80mm}
\includegraphics[width=75mm]{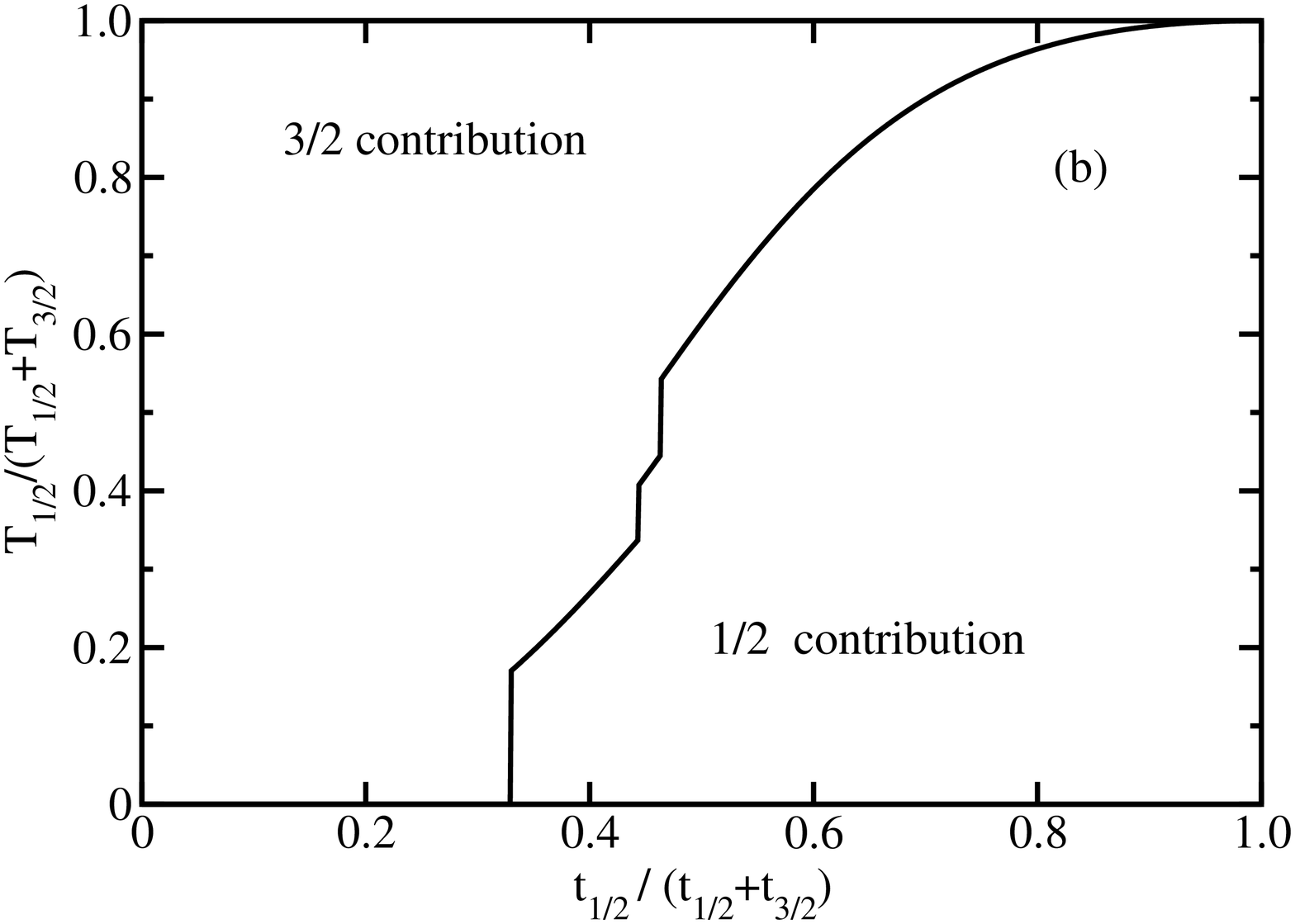}
\end{minipage}
\end{minipage}
\end{center}
\caption{Partial localization is reflected in the relative
contributions $T_{\alpha}$ (see text) of the various orbitals to the 
kinetic energy in the correlated ground-state for (a) t$_{5/2}$=0 and
(b) t$_{1/2}$=t$_{5/2}$. Dominant
hopping strongly suppresses the remaining contributions.}
\label{fig:PartialKineticEnergies}
\end{figure} 
Partial localization becomes clearly evident in the contributions of
the different j$_z$-channels to the gain in kinetic energy as shown in 
Figure \ref{fig:PartialKineticEnergies}. Whenever
one hopping parameter t$_\alpha$ dominates, i.~e.,~ 
t$_\alpha \gg$ t$_{\alpha'}$,  t$_{\alpha''}$  we find for the
corresponding ground-state expectation values
$T_{\alpha'}=
\frac{ \langle \Psi_0 |t_{\alpha'}
c_{\alpha'}^{\dagger}(1)c_{\alpha'}(2)|\Psi_0 \rangle}{\langle \Psi_0
|\sum_{\alpha} t_{\alpha}
c_{\alpha}^{\dagger}(1)c_{\alpha}(2)|\Psi_0 \rangle} \ll
\frac{t_{\alpha'}}{\sum_{\alpha}t_{\alpha}} \quad 
$
indicating that contributions of the smaller hopping matrix elements are
suppressed. 
 \section*{Acknowledgement} 
We would like to thank D. Efremov, N. Hasselmann, E. K. R. Runge, and
A. Yaresko for a number of helpful discussions.  We are grateful to 
S. R. Julian and G. G. Lonzarich for making 
their experimental data available to us prior to
publication. One of us (G.Z.) acknowledges support from the 
Niedersachsen-Israel Foundation. 

\end{document}